\documentclass[12pt]{iopart}
\usepackage{iopams}
\usepackage{setstack,amssymb}
\usepackage[dvipdfmx]{color}
\usepackage[dvipdfmx]{graphicx,color}
\usepackage{here}
\usepackage{ulem}

\newcommand{\dfracp}[2]{\frac{\partial #1}{\partial #2}}
\newcommand{\eqref}[1]{(\ref{#1})}
\newcommand{\wt}{\widetilde}
\newcommand{\wh}{\widehat}

\newcommand{\red}[1]{\textcolor{black}{#1}}
\newcommand{\blue}[1]{\textcolor{black}{#1}}

\newcommand{\magenta}[1]{\textcolor{black}{#1}}

\begin{document}

\title{Linear response theory \red{for} coupled phase oscillators with general coupling functions}
\author{Yu Terada$^{1,2}$ and Yoshiyuki Y. Yamaguchi$^{3}$}
\address{$^{1}$ Laboratory for Neural Computation and Adaptation, RIKEN Center for Brain Science, 2-1 Hirosawa, Wako, 351-0198 Saitama, Japan}
\address{$^{2}$ Department of Mathematical and Computing Science, Tokyo Institute of Technology, 152-8552 Tokyo, Japan}
\address{$^{3}$ Department of Applied Mathematics and Physics, Graduate School of Informatics, Kyoto University, Kyoto 606-8501, Japan}
\ead{yu.terada@riken.jp}

\begin{abstract}
  We develop a linear response theory \blue{by} computing the asymptotic value of the order parameter from the linearized equation of continuity around the nonsynchronized reference state \blue{using} the Laplace transform in time.
The proposed theory is applicable to a wide class of coupled phase oscillator systems and allows any coupling functions, any natural frequency distributions, any phase-lag parameters, and any values \blue{for the} time-delay parameter. 
This generality is in contrast to the limitation of the previous methods of the Ott--Antonsen ansatz and the self-consistent equation for an order parameter,  which are restricted to a model family whose coupling function consists of only a single sinusoidal function.
  The theory is verified by numerical simulations.
\end{abstract}

\maketitle

\section{Introduction}

Synchronization among rhythmic elements \blue{can be} ubiquitously observed in a wide variety of systems, whose \blue{numbers of elements range} from small to large \cite{strogatz-04,winfree-01,buzsaki-06,pikovsky-01}.
Since synchronization between two pendulum clocks hanging on the same wall was observed by Huygens, it \blue{has also been} found in \blue{many large systems,} such as flashing fireflies \cite{strogatz-04}, cardiac cells \cite{nitsan-16}, the circadian rhythm in mammals \cite{winfree-01}, and neuronal populations \cite{buzsaki-06,ermentrout-12}.
\blue{Synchronization} is often essential to biological functions, in large systems in particular, and we focus on large systems \blue{in this study}. 

\blue{Based on} mathematical modeling of a rhythmic element and the phase reduction theory \cite{kuramoto-03,nakao-16,kuramoto-19}, \blue{synchronization} is described by coupled phase-oscillator models.
For example, the Kuramoto model \cite{kuramoto-03}, a paradigmatic coupled phase-oscillator model, has  two types of states: 
\blue{a nonsynchronized state,} in which each oscillator rotates with its proper frequency called the natural frequency\red{,} and partially synchronized states\red{,} in which part of the oscillators rotate with the same effective frequency.
Strengthening the couplings provides \blue{a} synchronization transition from the nonsynchronized state to \blue{the} partially synchronized states, and the continuity of the transition is determined by the natural frequency distribution in the Kuramoto model \cite{kuramoto-03,kuramoto-75,strogatz-00,martens-barreto-strogatz-ott-so-antonsen-09,pazo-montbrio-09}.
We can thus reproduce macroscopic dynamics such as the continuous and discontinuous synchronization transitions by controlling \blue{the} microscopic details \magenta{that include} the coupling strength and the natural frequencies.

\blue{However, t}he single direction from a microscopic model to macroscopic dynamics does not provide a complete picture of the synchronization in real systems because we have \blue{often} no \blue{prior} knowledge of the microscopic character\blue{,} such as the coupling strength \blue{or} natural frequency distribution.
\blue{Additionally}, it is difficult to access directly the microscopic character.
\blue{This} motivates us to develop a theory \blue{that} extracts the microscopic character from \blue{a} macroscopic experiment.
\blue{One} potential candidate is the linear response theory;
a clear example of \blue{an} application is spectroscopy.

One strategy for obtaining the linear response formula in coupled oscillator models is to construct and analyze the self-consistent equation for the order parameter \blue{based on} knowledge of \blue{the} stationary states under an external force.
This strategy has been developed in models \blue{that have} a single sinusoidal coupling function\cite{sakaguchi-88,daido-15} and can reach the nonlinear regime beyond the linear response.
\blue{However, t}he construction of the self-consistent equation is not easily extended to general systems whose coupling functions consist of many harmonics \blue{of} sinusoidal functions because there are several stable stationary states for a given set of parameters \cite{komarov-pikovsky-13,komarov-pikovsky-14,li-etal-14}.

Another strategy relies on the Ott--Antonsen ansatz
\cite{ott-antonsen-08,ott-antonsen-09}.
This ansatz reduces the equation of continuity, which describes \blue{the} dynamics in the large population limit, to finite-dimensional ordinary differential equations.
See \blue{ref.} \cite{daido-15} for an application to the linear response.
\blue{However, t}he benefit of this reduction is limited in the single harmonic case again, and the natural frequency distribution must be rational, \blue{e.g.,} a Lorentzian, so that one can apply the residue theorem.

The above two strategies \blue{result in} difficulties when we consider a general coupling function, while coupling functions often consist of several harmonics in neuronal networks \cite{hansel-mato-meunier-93}, in electrochemical oscillators \cite{kiss-zhai-hudson-05,kiss-zhai-hudson-06}, and near a Hopf bifurcation \cite{kori-kuramoto-jain-kiss-hudson-14,ashwin-rodrigues-16}.
In addition to the generality of coupling functions, a desired linear response theory must \blue{provide} susceptibility for \blue{all} rotation frequencies of external forces from \blue{the} point of view of \blue{experimental measurements}.
Furthermore, the time delay in couplings, which \blue{drastically} changes \blue{its} synchrony \cite{yeung-99,lee-09}, must be incorporated because it is inevitable in many natural systems.
\blue{Thus,} we propose a linear response theory \blue{that is} applicable to systems with any coupling functions accompanied by a time delay, any natural frequency distributions, and any rotational frequencies of external forces.

\blue{Linear} response theory has been developed in statistical mechanics \cite{kubo-57,kubo-toda-hashitsume-85}, but computing a value of susceptibility is not easy
because we \blue{must} solve the equations of motion for an $N$-body system, which is nonintegrable in general.
In the class of globally coupled interactions, this difficulty has been overcome in the large population limit by using the Vlasov equation, which describes \blue{the} dynamics of \blue{a} one-body distribution function and is an analogy to the equation of continuity \cite{patelli-gupta-nardini-ruffo-12,ogawa-yamaguchi-12}.
Our approach has been inspired by this linear response theory.

The construction of this paper is as follows.
In Section \ref{sec:model}, we describe \blue{the} model which we analyze in this \blue{study}.
We \blue{describe the constructed} linear response theory based on the equation of continuity with a general coupling function and a time delay in Section \ref{sec:theory}. 
In Section \ref{sec:numerics}, we \blue{describe the} numerical simulations \blue{used} to validate the theory.
Section \ref{sec:summary} is devoted to \blue{the} summary and discussion.

\section{Model}\label{sec:model}

In this section\blue{,} we introduce our model, relevant quantities, and the large population limit.

\subsection{Settings}

We consider the system described by the ordinary differential equations 
\begin{equation}
  \label{eq:system}
  \frac{d\theta_{j}}{dt}(t)
  = \omega_{j} + \frac{1}{N} \sum_{k=1}^{N}
  \Gamma\left( \theta_{j}(t)-\theta_{k}(t-\tau) \right)
  + H \left( \theta_{j}(t),t \right),
\end{equation}
where $\theta_j(t)\in[0,2\pi)$ is the phase of the $j$th phase-oscillator, $\omega_{j}$ is its time-independent natural frequency, $\Gamma(\theta_j(t)-\theta_k(t-\tau))$ is the coupling function from oscillator $k$ to $j$, $\tau$ \blue{refers to} the time delay of \blue{the} couplings, and $H(\theta,t)$ is the external force.
We assume that function $H(\theta,t)$ has the form
\begin{equation}
  \label{eq:barH}
  H(\theta,t) = \Theta(t) \bar{H}(\theta-\omega_{\rm ex}t),
\end{equation}
where $\omega_{\rm ex}$ is the external frequency and $\Theta(t)$ is the Heviside step function\blue{.}
The external force is turned off for $t<0$ and is turned on at $t=0$.
To investigate the linear response, we assume that the external force is sufficiently small, $|H(\theta,t)|\ll 1$.

The natural frequencies follow the natural frequency distribution $g(\omega)$, which satisfies the normalization condition \red{of}
\begin{equation}
  \int_{-\infty}^{\infty}d\omega\,g(\omega)  = 1.
\end{equation}
We note that, for \blue{a} nonzero time delay \blue{of} $\tau>0$, system \eqref{eq:system} is not invariant under the rotating reference frame with frequency $\omega_{0}$ and \blue{a} shift \blue{in} $\bar{g}(\omega)=g(\omega-\omega_{0})$, \blue{whereas} this invariance holds for $\tau=0$.

Referring to the works by Kuramoto \cite{kuramoto-84} and Daido \cite{daido-92}, we expand the coupling function $\Gamma$ and the external force $H$ into the Fourier series as
\begin{equation}
  \label{eq:Gamma-sin}
  \Gamma(\theta) = - \sum_{n>0} K_{n} \sin(n\theta+\alpha_{n})
\end{equation}
and, from the form of the external force \eqref{eq:barH},
\begin{equation}
  \label{eq:H-sin}
  H(\theta,t) = - \Theta(t) \sum_{n>0} h_{n} \sin[n(\theta-\omega_{\rm ex}t)-\beta_n].
\end{equation}
\blue{Real} constants $K_{n}$, $\alpha_{n}$\blue{,} and $h_{n}$ \red{represent} the coupling strength, the phase-lag, and the amplitude of the external force for the Fourier mode $n$, respectively.
The real parameter $\beta_{n}$ determines the direction of the external force in the rotating reference frame with frequency $\omega_{\rm ex}$.

\subsection{Order parameters and susceptibilities}
To study \blue{the} synchrony of the oscillators, we introduce the order parameters
\begin{equation}
  \label{eq:order-parameters}
  z_{m}(t) = \frac{1}{N} \sum_{k=1}^{N} e^{im\theta_{k}(t)},
  \qquad
  (m=1,2,\cdots)
\end{equation}
which are called the Daido order parameters \cite{daido-92} and detect \blue{the} clusters whose phases are congruent modulo $2\pi/m$.
The first-order parameter\red{,} $z_{1}$\red{,} is equivalent to the Kuramoto order parameter \cite{kuramoto-84}.
Owing to the rotating external force $H(\theta,t)$, the order parameters also rotate with external frequency $\omega_{\rm ex}$.
Accordingly, the susceptibility tensor $\boldsymbol{\chi}(\omega_{\rm ex})=(\chi_{mn}(\omega_{\rm ex}))$ depending on the external frequency is asymptotically defined by
\begin{equation}
  \label{eq:definition-susceptibility}
  \chi_{mn}(\omega_{\rm ex}) e^{in\omega_{\rm ex}t}
  = \frac{\partial z_{m}}{\partial h_{n}}\Bigg|_{\mathbf{h}\to \mathbf{0}}(t) ,
  \quad
  \mathrm{as}
  \quad
  t\to\infty.
\end{equation}
In other words, $\chi_{mn}(\omega_{\rm ex})$ is defined as the rate of change in the $m$th Daido order parameter induced by the $n$th Fourier mode of the external force
  in the rotating reference frame with frequency $\omega_{\rm ex}$.
We aim to obtain the susceptibility $\chi_{mn}(\omega_{\rm ex})$ \blue{by} taking the large population limit in a coupled oscillator system \blue{that has} a general coupling function.

\subsection{Large population limit}
Using the Daido order parameters \eqref{eq:order-parameters}, we have a simple expression \blue{for} the equations of motion \eqref{eq:system} as
\begin{equation}
  \label{eq:system-MF}
  \frac{d\theta_{j}}{dt}(t)
  = \omega_{j}
  + \sum_{n=-\infty}^{\infty} 
  \left[ \wt{\Gamma}_{n} z_{-n}(t-\tau) + \wt{H}_{n}(t) \right]
  e^{in\theta_{j}},
\end{equation}
where $\wt{\Gamma}_{n}$ and $\wt{H}_{n}(t)$ are the Fourier components of the coupling function $\Gamma(\theta)$ and the external force $H(\theta,t)$, which are defined by
\begin{equation}
  \label{eq:Gamma-Fourier}
  \Gamma(\theta) = \sum_{n=-\infty}^{\infty} e^{in\theta}~ \wt{\Gamma}_{n}
\end{equation}
and
\begin{equation}
  \label{eq:H-Fourier}
  H(\theta,t) = \sum_{n=-\infty}^{\infty} e^{in\theta}~ \wt{H}_{n}(t).
\end{equation}
The present forms of $\Gamma$ \eqref{eq:Gamma-sin} and $H$ \eqref{eq:H-sin} give
\begin{equation}
  \wt{\Gamma}_{n} = \frac{i}{2} K_{n} e^{i\alpha_{n}}, \qquad
  \wt{\Gamma}_{-n} = \wt{\Gamma}_{n}^{\ast}, \qquad (n>0),
\end{equation}
and
\begin{equation}
  \wt{H}_{n}(t) = \frac{i}{2} h_{n} \Theta(t) e^{-i(n\omega_{\rm ex}t+\beta_n)}, \qquad
  \wt{H}_{-n}(t) = \wt{H}_{n}^{\ast}(t), \qquad (n>0),
  \label{eq:H_forms}
\end{equation}
where $\wt{\Gamma}_{n}^{\ast}$ is the complex conjugate of $\wt{\Gamma}_{n}$\red{,} for instance, and we define $\wt{\Gamma}_0=\wt{H}_0=0$.

In the large population limit \blue{of} $N\to\infty$, the dynamics of the system \eqref{eq:system-MF} \blue{are} described by the equation of continuity
\begin{equation}
  \label{eq:eq-continuity}
  \frac{\partial F}{\partial t}
  + \frac{\partial }{\partial \theta}\left( V F \right) = 0,
\end{equation}
where $F(\theta,\omega,t)$ is the probability density function \blue{that satisfies} the normalization condition
\begin{equation}
  \label{eq:F-normalization}
  \int_{0}^{2\pi} d\theta \int_{-\infty}^{\infty} d\omega~ F(\theta,\omega,t) = 1,
\end{equation}
and velocity $V$ is given by
\begin{equation}
  V(\theta,\omega,t) = \omega + \sum_{n=-\infty}^{\infty} e^{in\theta}~
  \left[ \wt{\Gamma}_{n} z_{-n}(t-\tau) + \wt{H}_{n}(t) \right].
\end{equation}
The Daido order parameters are expressed as
\begin{equation}
  z_{m}(t) = \int_{0}^{2\pi} d\theta \int_{-\infty}^{\infty} d\omega~
  e^{im\theta} F(\theta,\omega,t),\qquad
  (m=1,2,\cdots).
\end{equation}

\section{Linear response theory}\label{sec:theory}
We derive the susceptibility in the nonsynchronized state by solving the linearized equation of continuity around the nonsynchronized state.
The solution is obtained by performing the Fourier transform in the phase variable and the Laplace transform in the time variable.

\subsection{Linearization around the nonsynchronized state}

As the reference state for $t<0$, we consider the nonsynchronized state
\begin{equation}
  F(\theta,\omega,t) = F_{0}(\omega),
  \quad
  \mathrm{for}
  \quad
  t< 0,
\end{equation}
which is the trivial stationary solution to the equation of continuity \eqref{eq:eq-continuity}.
The normalization condition \eqref{eq:F-normalization} gives
\begin{equation}
  F_{0}(\omega) = \frac{g(\omega)}{2\pi}.
\end{equation}
The small external force $H(\theta,t)$ \blue{that starts} at $t=0$ perturbs the state from $F_{0}$ to
\begin{equation}
  \label{eq:F-expand}
  F(\theta,\omega,t)
  = F_{0}(\omega) + f(\theta,\omega,t)
  \quad
  \mathrm{for}
  \quad
  t\geq 0,
\end{equation}
where $|f(\theta,\omega,t)|\ll1$ is assumed according to $|H(\theta,t)|\ll 1$.
The continuity of $F(\theta,\omega,t)$ at $t=0$ implies \blue{that} the initial condition of $f(\theta,\omega,t)$ \blue{is}
\begin{equation}
  \label{eq:f-initial}
  f(\theta,\omega,0) = 0.
\end{equation}

The Daido order parameters \blue{receive} no contribution from the nonsynchronized reference state\red{,} $F_{0}(\omega)$\red{,} and they have the same order with perturbation $f$ as
\begin{equation}
  z_{n}(t) = \int_{0}^{2\pi} d\theta \int_{-\infty}^{\infty} d\omega~
  e^{in\theta} f(\theta,\omega,t).
\end{equation}
This \red{ordering} \blue{produces} the linearized equation of continuity as
\begin{eqnarray}
  \dfracp{f}{t}(\theta,\omega,t)
  + \omega \dfracp{f}{\theta}(\theta,\omega,t)
  + F_{0}(\omega) \left[
    \dfracp{v}{\theta}(\theta,t,\tau)
    + \dfracp{H}{\theta}(\theta,t)
  \right]
  = 0,
  \label{eq:f1-equation}
\end{eqnarray}
where we defined
\begin{equation}
  v(\theta,t,\tau) = \sum_{n=-\infty}^{\infty} e^{in\theta}
  ~ \wt{\Gamma}_{n} z_{-n}(t-\tau) .
\end{equation}
The linearized equation \eqref{eq:f1-equation} is valid under a weak external force.

\subsection{Fourier--Laplace transforms}
To solve the linearized equation \eqref{eq:f1-equation}, we employ the Fourier and Laplace analyses.
Substituting the Fourier series expansion of $f$,
\begin{equation}
  f(\theta,\omega,t) = \sum_{n=-\infty}^{\infty} e^{in\theta} \wt{f}_{n}(\omega,t),
\end{equation}
and of $H$ \eqref{eq:H-Fourier} into the linearized equation \eqref{eq:f1-equation}, we \blue{find} the Fourier-expanded expression for each $n$ as
\begin{equation}
  \label{eq:iinearized-eq-Fourier}
  \fl
  \dfracp{\wt{f}_{-n}}{t}(\omega,t)
  - in\omega \wt{f}_{-n}(\omega,t)
  - in F_{0}(\omega) \left[
    \wt{\Gamma}_{-n} z_{n}(t-\tau) + \wt{H}_{-n}(t)
  \right] = 0.
\end{equation}
Further, we perform the Laplace transform which is defined for $\varphi(t)$ \red{to be}
\begin{equation}
  \label{eq:Laplace-transform-def}
  \wh{\varphi}(s) = \int_{0}^{\infty}dt\,\varphi(t) e^{-st}
  \quad
  \mathrm{for}\quad
  {\rm Re}(s)>0,
\end{equation}
where the condition ${\rm Re}(s)>0$ is assumed \blue{to ensure} the convergence of the integral.
The Laplace transform modifies  equation \eqref{eq:iinearized-eq-Fourier} as
\begin{equation}
  \label{eq:f-equation-Laplace}
  \fl
  (s-in\omega) \wh{f}_{-n}(\omega,s)
  = in F_{0}(\omega)
  \left[
    \wt{\Gamma}_{-n} e^{-s\tau} \wh{z}_{n}(s) + \wh{H}_{-n}(s) 
  \right]
  + \wt{f}_{-n}(\omega,0),
\end{equation}
where we used $z_n(t)=0$ for $t<0$.
We note that the last term \blue{disappears} owing to the initial condition \eqref{eq:f-initial}\blue{; however} this term is kept tentatively \blue{to discuss} the stability of the reference state\red{,} $F_{0}(\omega)$.

Both sides of Eq. \eqref{eq:f-equation-Laplace} contain $\wh{f}_{-n}$, and $\wh{f}_{-n}$ is associated with the Laplace transform of the order parameter, $\wh{z}_{n}(s)$, as
\begin{equation}
  \wh{z}_{n}(s) = 2\pi \int_{-\infty}^{\infty}d\omega\, \wh{f}_{-n}(\omega,s).
\end{equation}
Using this relation, we solve Eq. \eqref{eq:f-equation-Laplace} with respect to $\wh{z}_{n}(s)$.
\blue{By multiplying} \eqref{eq:f-equation-Laplace} by $2\pi(s-in\omega)^{-1}$ and integrating over $\omega$, we have
\begin{equation}
  \fl
  \wh{z}_{n}(s) = 
  in\left[
    \wt{\Gamma}_{-n} e^{-s\tau} \wh{z}_{n}(s)
    + \wh{H}_{-n}(s)
  \right]
  \int_{-\infty}^{\infty}d\omega\frac{g(\omega)}{s-in\omega} 
  + 2 \pi \int_{-\infty}^{\infty}d\omega\frac{\wt{f}_{-n}(\omega,0)}{s-in\omega}.
\end{equation}
\red{By d}efining the functions
\begin{equation}
  \label{eq:In}
  I_{n}(s) = in\int_{-\infty}^{\infty}d\omega\frac{g(\omega)}{s-in\omega},
\end{equation}
and
\begin{equation}
  \label{eq:Lambdan}
  G_{n}(s)
  = 1 - \frac{K_{n}}{2i} e^{-i\alpha_{n}} e^{-s\tau} I_{n}(s),
\end{equation}
the Laplace transform of the order parameter $z_{n}(t)$ is expressed as
\begin{equation}
  \label{eq:response-formula}
  \wh{z}_{n}(s) = \frac{1}{G_{n}(s)} \left[
    I_{n}(s) \wh{H}_{-n}(s)
    + 2 \pi \int_{-\infty}^{\infty}d\omega\frac{\wt{f}_{-n}(\omega,0)}{s-in\omega}
  \right].
\end{equation}
The temporal evolution of $z_{n}(t)$ is obtained by performing the inverse Laplace transform.

The inverse Laplace transform of function $\wh{\varphi}(s)$ is defined by
\begin{equation}
  \varphi(t) = \frac{1}{2\pi i} \int_{{\rm Br}}ds\,\wh{\varphi}(s) e^{st},
\end{equation}
where the integral is performed along the Bromwich contour which lies on the right-hand side of any singularities of $\wh{\varphi}(s)$ on the complex $s$ plane.
If $\wh{\varphi}(s)$ has a simple pole at $s=s_{0}\in\mathbb{C}$, the inverse Laplace transform produces \blue{a} mode proportional to $\exp(s_{0}t)$ \blue{upon} picking up this pole in the residue theorem. 
As a result, a pole of ${\rm Re}(s_{0})<0$ corresponds to a stable mode, and a pole of ${\rm Re}(s_{0})>0$ \blue{corresponds} to an unstable mode.

We note that integral $I_{n}(s)$, and function $G_{n}(s)$ are defined on the right-half plane \blue{of} ${\rm Re}(s)>0$ of the complex $s$ plane following the definition of the Laplace transform \eqref{eq:Laplace-transform-def}.
To apply the discussion above to $\wh{z}_{n}(s)$, we need to perform the analytic continuation of $I_{n}(s)$.
See \ref{sec:continuation} for \blue{this} continuation.

The stability of the nonsynchronized state $F_{0}(\omega)$ is examined by turning off the external force, $H(\theta,t)=0$, and is determined by the roots of $G_{n}(s)$.
\blue{A} demonstration of the stability analysis is \blue{provided} in \ref{sec:Nyquist} \blue{with the use of \red{\sout{the}} Nyquist diagrams}.
Assuming the stability of the nonsynchronized state $F_{0}(\omega)$ through the rest of \blue{this study}, we compute the susceptibility from \eqref{eq:response-formula}, whose last term vanishes because of the initial condition \eqref{eq:f-initial}.

\subsection{Susceptibilities}
\label{sec:susceptibilities}
The susceptibility $\chi_{mn}(\omega_{\rm ex})$ is defined in the limit \blue{of} $t\to\infty$ as in \eqref{eq:definition-susceptibility}, and the asymptotic temporal evolution of $z_{n}(t)$ is obtained \blue{by} considering the singularities of \eqref{eq:response-formula}.
The stability assumption of $F_{0}$ implies that all the roots of $G_{n}(s)$ are on the left-half plane ${\rm Re}(s)<0$.
Consequently, the asymptotic behavior of the order parameters is determined by the residues of the poles on the imaginary axis\blue{,} which come from the external force as, from \eqref{eq:H_forms},
\begin{equation}
  \wh{H}_{n}(s) = - \frac{1}{2i} \frac{h_{n}e^{-i\beta_n}}{s+in\omega_{\rm ex}},
  \quad
  \wh{H}_{-n}(s) = \frac{1}{2i} \frac{h_{n}e^{i\beta_n}}{s-in\omega_{\rm ex}},
  \quad
  (n>0).
\end{equation}
The residue theorem gives
\begin{equation}
  z_{n}(t)
  \to \frac{I_{n}(in\omega_{\rm ex})}{2iG_{n}(in\omega_{\rm ex})}
  e^{i(n\omega_{\rm ex}t+\beta_n)} h_{n},
   \quad
  \blue{\mathrm{as}
   \quad
   t\to\infty}
  \quad
  (n>0)
\end{equation}
and
\begin{equation}
  z_{-n}(t)
  \to \frac{-I_{-n}(-in\omega_{\rm ex})}{2iG_{-n}(-in\omega_{\rm ex})}
  e^{-i(n\omega_{\rm ex}t+\beta_n)} h_{n},
  \quad
   \blue{\mathrm{as}
   \quad
   t\to\infty}
  \quad
  (n>0),
\end{equation}
where, for $n>0$,
\begin{equation}
  I_{n}(in\omega_{\rm ex})
  = - \left[ J(\omega_{\rm ex}) - i\pi g(\omega_{\rm ex}) \right],
  \quad
  I_{-n}(-in\omega_{\rm ex})
  = [I_{n}(in\omega)]^{\ast},
\end{equation}
and
\begin{equation}
  \label{eq:definition_J}
  J(\omega_{\rm ex})
  = {\rm PV} \int_{-\infty}^{\infty}d\omega\frac{g(\omega)}{\omega-\omega_{\rm ex}}.
\end{equation}
PV represents the Cauchy principal value.
>From the definition \eqref{eq:definition-susceptibility}, the susceptibility tensor $\boldsymbol{\chi}(\omega_{\rm ex})=(\chi_{mn}(\omega_{\rm ex}))$ is diagonal as
\begin{equation}
  \chi_{mn}(\omega_{\rm ex}) = \chi_{n}(\omega_{\rm ex}) \delta_{mn},
\end{equation}
and, for $n>0$,
\begin{equation}
  \label{eq:chin}
  \chi_{n}(\omega_{\rm ex})
  = \frac{\pi g(\omega_{\rm ex}) + iJ(\omega_{\rm ex})}
  {2 - K_{n} e^{-i(\alpha_{n}+n\omega_{\rm ex}\tau)} [ \pi g(\omega_{\rm ex}) + iJ(\omega_{\rm ex}) ]}e^{i\beta_n},
\end{equation}
and
\begin{equation}
  \chi_{-n}(\omega_{\rm ex}) = \chi_{n}^{\ast}(\omega_{\rm ex}).
\end{equation}
The susceptibility \eqref{eq:chin} is the main result of this \blue{study}. 
\blue{This} main result \eqref{eq:chin} is an extension of the susceptibility \cite{sakaguchi-88} obtained in the Kuramoto model \cite{kuramoto-03,kuramoto-75}\blue{,} whose coupling function is $\Gamma(\theta)=-\sin\theta$ without the time-delay.
We \blue{provide} three remarks derived from \blue{this} main result \eqref{eq:chin}.

The first remark is on \blue{the} divergence of the susceptibility.
The divergence appears if the complex denominator of the susceptibility \eqref{eq:chin} vanishes.
For instance, the divergence is observed at $K_{n,{\rm c}}=2/[\pi g(0)]$ if the model has three types of symmetry: \blue{a} zero phase lag $\alpha_{n}=0$, \blue{a} zero time delay $\tau=0$, and \blue{a} symmetric natural frequency $g(\omega)$, where the satisfaction of the last symmetry gives $J(0)=0$.
\blue{In contrast,} the divergence does not appear for any value of $K_{n}$ unless we tune the external frequency $\omega_{\rm ex}$ \blue{so that the imaginary part of the denominator vanishes}\red{\sout{,}} if one of the three types of symmetry is broken.
The absence of divergence is demonstrated in \ref{eq:Kuramoto-model} in the Kuramoto model with an asymmetric natural frequency.
In the next section, we \blue{discuss our examination of} the theory with the tuned $\omega_{\rm ex}$ to observe the divergence because the divergence is a representative phenomenon at the critical point, and \blue{it} provides a deep numerical examination: the divergence expands the discrepancy of susceptibility between the theoretically predicted value and the numerically observed value \blue{that is} perturbed by breaking \blue{the} assumption of the large population limit or the zero external force limit.

The second remark is on the critical exponent $\gamma$, which is defined as
\begin{equation}
  \chi \propto |K-K_{\rm c}|^{-\gamma}
\end{equation}
around the critical coupling strength $K_{\rm c}$ \cite{nishimori-10}.
\blue{We} choose $\omega_{\rm ex}$ to observe the divergence.
The denominator of the susceptibility \eqref{eq:chin} depends on the coupling constant linearly, and the critical exponent $\gamma$ should be unity: $\gamma=1$.
This universality of $\gamma$ gives a sharp contrast with the dependence of the critical exponent $\beta$ on the coupling function: the single harmonic case has $\beta=1/2$ \cite{kuramoto-84} while $\beta=1$ for the general case \cite{daido-94,crawford-95}.
The critical exponent $\beta$ is defined here, with absence of the external force, by
\begin{equation}
  \label{eq:beta-definition}
  z_{1} \propto (K-K_{\rm c})^{\beta}
\end{equation}
above and around the critical point.

The last remark is \blue{regarding} the roles of the phase lag and the time delay.
The phase lag $\alpha_{n}$ and the time delay $\tau$ are included only in the exponential factor of the denominator and play a similar role in the susceptibility \eqref{eq:chin}.
In fact, the phase lag can be eliminated from the time-delayed Kuramoto model if the external frequency\red{,} $\omega_{\rm ex}$\red{,} is fixed
\cite{yeung-99}.
However, the time delay couples \blue{to} $\omega_{\rm ex}$, and, from experimental observation of the susceptibility, we can identify which factor is included in the system by varying $\omega_{\rm ex}$.

\section{Numerical tests}
\label{sec:numerics}
To verify our theoretical results, we perform numerical simulations, where we also shed light on \blue{the} difference between the roles of the phase lag $\alpha_{n}$ and the time delay $\tau$.
We first examine the linear response in the Sakaguchi--Kuramoto model, which contains the phase lag but not the time delay in a single sinusoidal coupling function.
The single harmonics permits us to apply the Ott--Antonsen reduction, which eliminates finite-size fluctuations and is useful \blue{for examining} the theory.
The second model is the time-delayed Daido--Kuramoto model, which has multi-harmonics in the coupling functions.
This model highlights \blue{the} usefulness of the proposed theory for general systems.
In the above two models, we use the Lorentz distribution
\begin{equation}
  g(\omega) = \frac{\gamma}{\pi}\frac{1}{(\omega-\Omega)^{2}+\gamma^{2}}
  \label{eq:lorentz}
\end{equation}
as the natural frequency distribution, although the proposed theory can be applied to other distributions.
Note that the time delay breaks the invariance of \blue{the} system with respect to the Galilei transform of $\theta$ \blue{and} \magenta{hence} the rotation of the reference frame and $\Omega$ \red{cannot} be removed by shifting $\omega$ if the coupling function has a nonzero time delay\red{,} $\tau$.
The principal value\red{,} $J(\omega_{\rm ex})$ \eqref{eq:definition_J}, can be explicitly computed for the Lorentzian \eqref{eq:lorentz}\blue{,} as shown in \ref{sec:lorentz_case}.
  
We adopt the 4th-order Runge--Kutta method to integrate the original equations \eqref{eq:system} and the reduced equations\magenta{, where the time length is $T=1000$ with \blue{a} discrete time size $dt=0.05$}.
Values of the order parameters are evaluated by averaging over the
\red{time interval of $[T/2,T]$.}

\subsection{The Sakaguchi--Kuramoto model}
\label{sec:sk_model}
The Sakaguchi--Kuramoto model is recovered by setting the coupling function\red{,} $\Gamma(\theta)$\red{, to be}
\begin{equation}
  \label{eq:SKmodel}
  \Gamma(\theta) = - K \sin(\theta+\alpha),
\end{equation}
without the time-delay, where $0<\alpha<\pi$.
The first susceptibility is read as
\begin{equation}
  \label{eq:chi1-SK}
  \chi_{1}(\omega_{\rm ex})
  = \frac{\pi g(\omega_{\rm ex}) + i J(\omega_{\rm ex})}
  {2-K e^{-i\alpha} [\pi g(\omega_{\rm ex}) + i J(\omega_{\rm ex})] }.
\end{equation}
We choose the frequency of the external force\blue{, $\omega_{\rm ex}$,} so that the system has the divergence of $\chi_{1}$ at the critical point.
The divergence appears when the denominator of \eqref{eq:chi1-SK} vanishes.
The imaginary part \blue{provides} the condition for $\omega_{\rm ex}$ as
\begin{equation}
  \label{eq:divergence-SK}
  J(\omega_{\rm ex}) \cos\alpha - \pi g(\omega_{\rm ex}) \sin\alpha = 0,
\end{equation}
and the real part determines the critical point\red{,} $K_{\rm c}$\red{,} as
\begin{equation}
  \label{eq:Kc-SK}
  K_{\rm c} = \frac{2\cos\alpha}{\pi g(\omega_{\rm ex})}.
\end{equation}

\begin{figure}[h]
  \centering
  \includegraphics[scale=0.65]{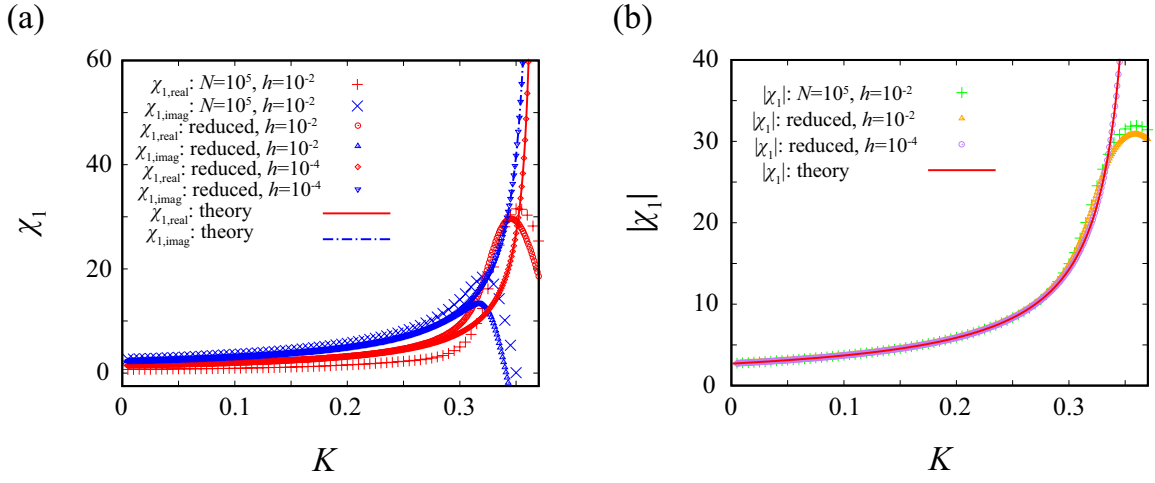}
  \caption{(a) Susceptibility and (b) its absolute value in the Sakaguchi--Kuramoto model \eqref{eq:SKmodel}.
    The natural frequency distribution $g(\omega)$ is the Lorentzian \eqref{eq:lorentz} with $(\gamma,\Omega)=(0.1,3)$.
    The phase-lag parameter is set \red{to be} $\alpha=1$ and the frequency of the external force is correspondingly chosen \red{to be} $\omega_{\rm ex}=2.844259$ to satisfy the divergence condition \eqref{eq:divergence-SK}.
    The critical point is $K_{\rm c}\simeq 0.370163$, which corresponds to the right boundaries of the panels.
    The numerical simulations are conducted with $N=10^{5}$ and $h=10^{-2}$ in $N$-body system, and with $h=10^{-2}$ and $h^{-4}$ in the reduced system.
    }
  \label{fig:susceptibility_sk_omega_ex}
\end{figure}

For \blue{a} fixed phase-lag parameter $\alpha=1$, susceptibility $\chi_{1}$ obtained theoretically, \eqref{eq:chi1-SK}, is shown in \blue{Figure} \ref{fig:susceptibility_sk_omega_ex} with numerical results in the $N$-body system and in the reduced system derived by the Ott--Antonsen ansatz.
The reduced system\red{, corresponding to the limit $N\to\infty$,} is in good agreement with the theory for a sufficiently small external force, $h=10^{-4}$, \blue{whereas} the agreement is not perfect \red{for $h=10^{-2}$ which}
  is not sufficiently small for imitating the limit of \red{$h\to 0$ taken in the definition of susceptibility, \eqref{eq:definition-susceptibility}.} 
\red{This observation concludes that the discrepancy between the theory and numerics} \magenta{results} \red{from the nonlinearity of the response
  with respect to $h$ rather than the finite-size fluctuation
  in the $N$-body system.}
  \magenta{We have two remarks in order.}
  \red{First, to observe the linear response clearly,
  $h$ in the $N$-body system must be larger than $O(1/\sqrt{N})$,
  which is the expected level of} \magenta{the} \red{finite-size fluctuation of the order parameter.
  Thus, we must use a larger $N$ to use a smaller $h$.
  Second, the nonlinearity inhibits the divergence at the critical point
  because the divergence results from the linear response.
  The nonlinearity, therefore, enhances the discrepancy around the critical point as observed for $h=10^{-2}$ in Figure \ref{fig:susceptibility_sk_omega_ex}.
}

\subsection{The time-delayed Daido--Kuramoto model}
\label{sec:daido_kuramoto}
The time-delayed Daido--Kuramoto model has the coupling function \blue{of}
\begin{equation}
  \Gamma(\theta) = - \sum_{n=1}^{M} K_{n} \sin\theta
\end{equation}
with  $M>1$ and \blue{the} nonzero time-delay parameter $\tau$ in \eqref{eq:system}.
The susceptibilities are explicitly written as
\begin{equation}
  \chi_{n}(\omega_{\rm ex})
  = \frac{\pi g(\omega_{\rm ex}) + i J(\omega_{\rm ex})}
  {2-K_ne^{-in\omega_{\rm ex}\tau} [\pi g(\omega_{\rm ex}) + i J(\omega_{\rm ex})] },
  \quad
  \mathrm{for}
  \quad
  1\le n\le M.
  \label{eq:chi_dk_model}
\end{equation}
This susceptibility for $n=1$ is obtained by replacing $\alpha$ with $\omega_{\rm ex}\tau$ in the susceptibility \eqref{eq:chi1-SK} of the Sakaguchi--Kuramoto model.
We adopt $M=2$ in \blue{the} numerical simulations for simplicity, although the theory is applicable to \blue{any} arbitrary $M$.
The coupling constant of the second Fourier mode, $K_{2}$, will be sufficiently small positive number so that the nonsynchronized state becomes unstable in the first Fourier mode.
We set the time-delay parameter as $\tau=2$ here and choose the value of $\omega_{\rm ex}$ \blue{to} satisfy the divergence condition.

It is worth commenting on the continuity of the synchronization transition.
\blue{First,} the theory is applicable in the nonsynchronized state even if the transition is discontinuous.
One advantage of the continuous transition is that numerical examinations can be \blue{robustly} performed, while the discontinuous transition, owing to the bistability, may \blue{accompany} a jump of responses around the critical point for a finite external force.
The parameter set, $\gamma=0.1,\Omega=3$, and $\tau=2$, gives the continuous transition in the delayed Kuramoto model \cite{yeung-99}, but it is not obvious if the time-delayed Daido--Kuramoto model also exhibits the continuous transition.
For instance, the multi-harmonic coupling function with $M=2$ \blue{results in} the discontinuous transition if $K_{2}>0$ and $\tau=0$ \cite{chiba-nishikawa-11}.
Our numerical computation, however, implies that the transition is continuous with the given parameter set.

\begin{figure}
  \centering
  \includegraphics[scale=0.65]{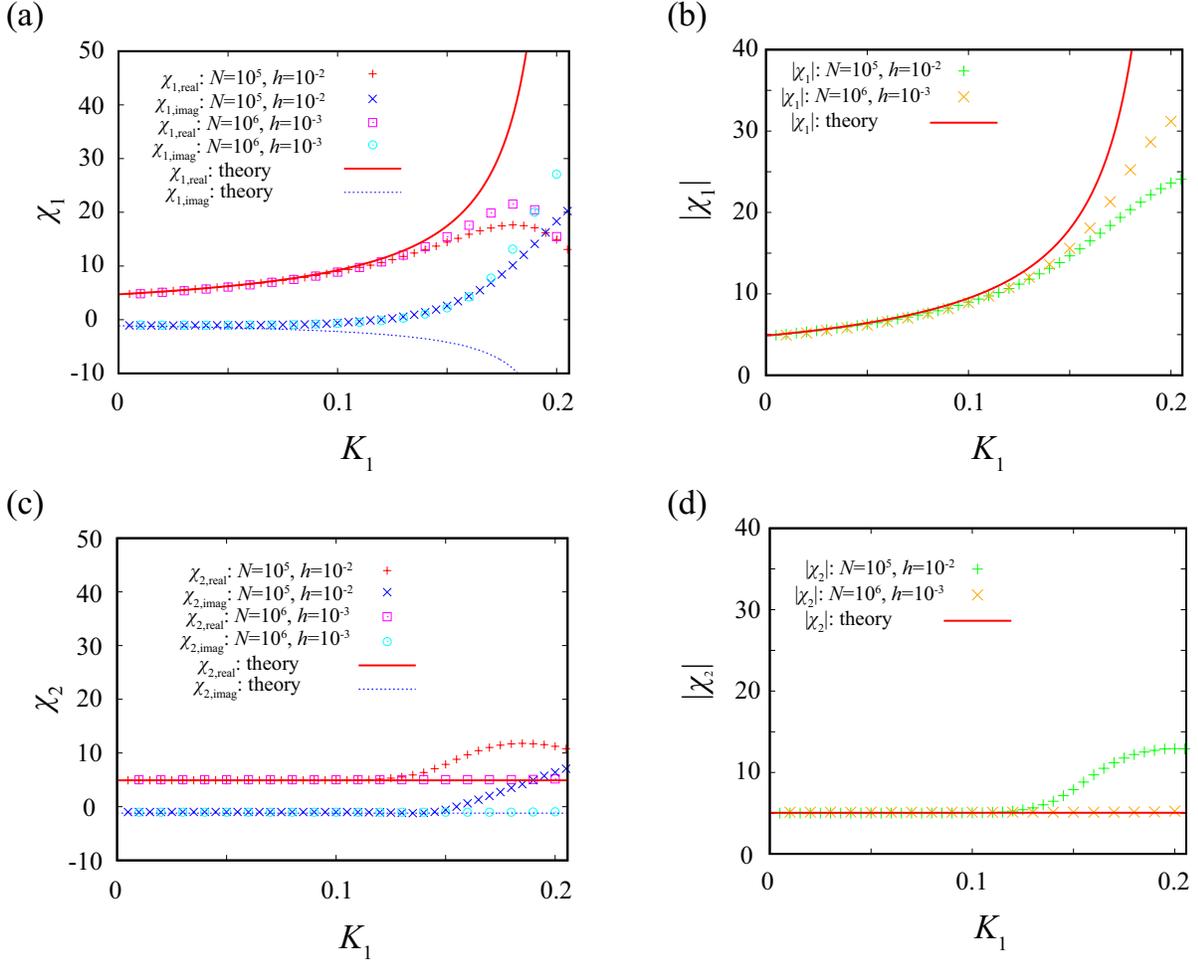}
  \caption{(a,c) Susceptibilities and (b,d) their absolute values in the time-delayed Daido--Kuramoto model with $M=2$.
  The natural frequency distribution\red{,} $g(\omega)$\red{,} is Lorentzian \eqref{eq:lorentz} with $(\gamma,\Omega)=(0.1,3)$.
  The delay parameter is set \red{to be} $\tau=2$ and  the frequency of the external force is determined \red{to be} $\omega_{\rm ex}=3.023969$ to observe the divergence of $\chi_1$.
  The critical point is $K_{1,\rm c}\simeq 0.205665$, which is the right boundaries of the panels.
  We set $K_2=0.01$ \blue{to keep mode $2$ stable}.
  The numerical simulations are conducted with $(N,h)=(10^{5},10^{-2})$ and $(10^{6},10^{-3})$.
  The susceptibility $\chi_{2}$ must be constant of $K_{1}$ because the susceptibility $\chi_{2}$ does not depend on $K_{1}$.}
  \label{fig:daido-kuramoto}
\end{figure}

The Ott--Antonsen reduction is not applicable to the Daido--Kuramoto model, and we show \blue{the} results \blue{for} only $N$-body simulations \blue{in Figure \ref{fig:daido-kuramoto}}.
The numerical results do not perfectly agree with the theoretical prediction, but they \blue{approach} the theoretical curves as the external force \blue{becomes} small.
This tendency of discrepancies is very similar to the ones observed in the Sakaguchi--Kuramoto model, in which the validity of the theory has been confirmed with the aid of the reduced system.
\blue{Thus, we} conclude that the theory is \blue{also} valid in the time-delayed Daido--Kuramoto model.

\subsection{Distinction between the phase lag and the time delay}
\begin{figure}[ht]
  \centering
  \includegraphics[scale=0.3]{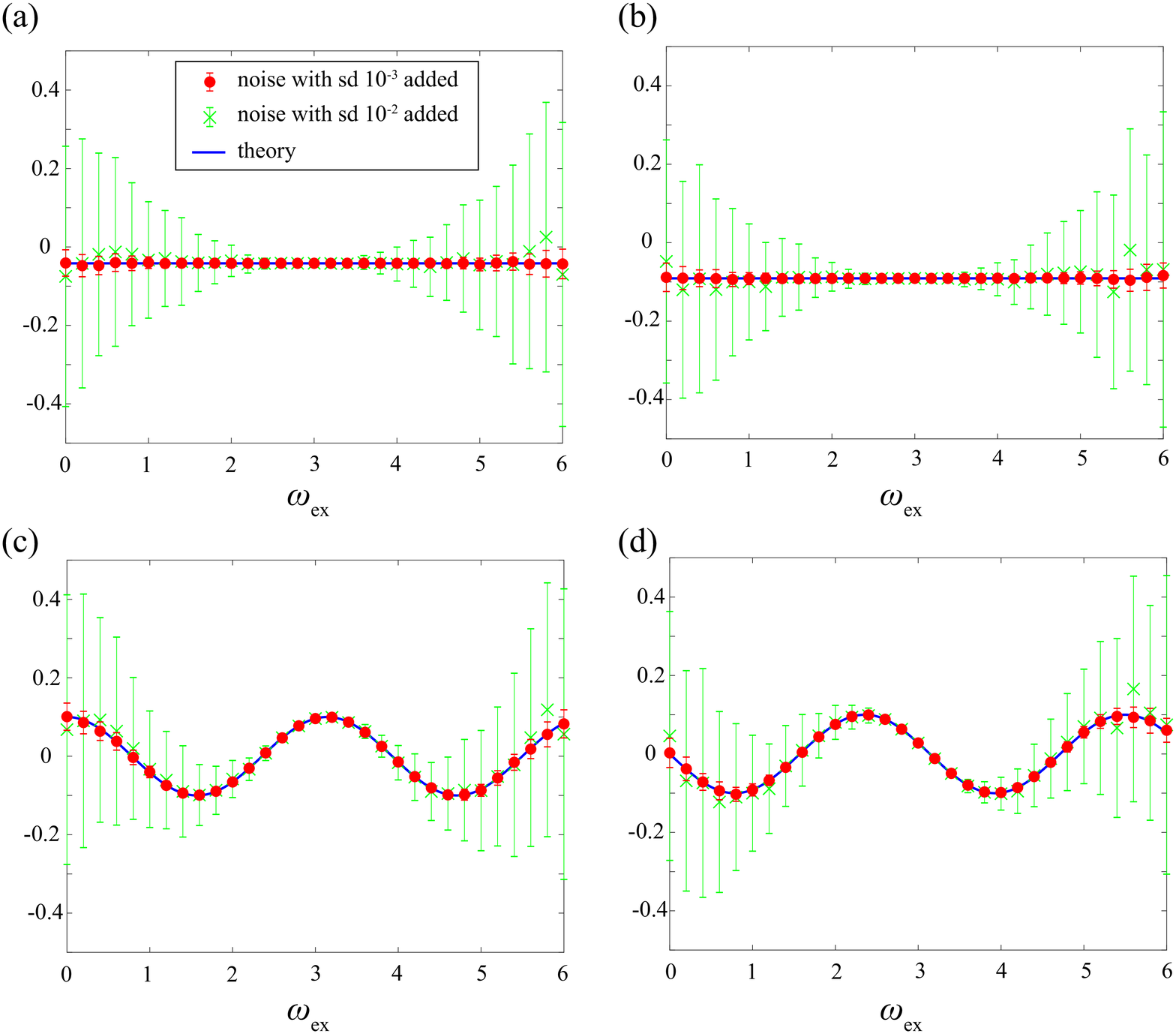}
  \caption{(a,c) Real and (b,d) imaginary parts of the right-hand side of \eqref{eq:tau_dependency2}, under replacing the theoretical susceptibility $\chi_{1}(\omega_{\rm ex})$ with a noisy one $\chi_{1}^{(l)}(\omega_{\rm ex})$.
     (a,b) $(\tau,\alpha)=(0,2)$ and (c,d) $(\tau,\alpha)=(2,0)$.
   \blue{The points} represent the means over $100$ realizations, and the error bars the standard deviations.
   Noises follow the zero-mean normal distributions and their standard deviations $\sigma$ are $\sigma=10^{-3}$ (red circles) and $10^{-2}$ (green crosses).
   The blue solid lines are the theoretical predictions.
   The natural frequency $g(\omega)$ is the Lorentzian \eqref{eq:lorentz} with $(\gamma,\Omega)=(0.1,3)$, and the coupling strength is set \red{to be} $K_{1}=0.1$.}
  \label{fig:KE-test}
\end{figure}

Finally, we illustrate that phase lag $\alpha$ plays the same role as time delay $\tau$ for a fixed $\omega_{\rm ex}$ but a different role for a varying $\omega_{\rm ex}$.
For simplicity, we concentrate on the time-delayed Sakaguchi--Kuramoto model and assume that the coupling strength $K_{1}$ and the natural frequency distribution\red{,} $g(\omega)$\red{,} are known.
This assumption implies that, in a rewritten form of \eqref{eq:chin},
\begin{equation}
  \label{eq:tau_dependency2}
  K_{1} e^{-i(\alpha_{1}+\omega_{\rm ex}\tau)}
  = \frac{2}{\pi g(\omega_{\rm ex}) +i J(\omega_{\rm ex})}
  - \frac{1}{\chi_{1}(\omega_{\rm ex})},
\end{equation}
the first term of the most right-hand side is known\blue{,} but the second term must be pointwisely observed in \blue{an} experiment with inevitable errors.
We imitate the errors by adding noise $\xi^{(l)}(\omega_{\rm ex})$ to the theoretical value \red{of} $\chi_{1}(\omega_{\rm ex})$ at each observing point $\omega_{\rm ex}$: $\chi_{1}^{(l)}(\omega_{\rm ex})=\chi_{1}(\omega_{\rm ex}) + \sigma \xi^{(l)}(\omega_{\rm ex})$.
The random value\red{,} $\xi^{(l)}(\omega_{\rm ex})$\red{,} is independently drawn from the standard normal distribution, and superscript $l$ \magenta{denotes the index of} realization, \red{in other words,} virtual observation.
In \blue{Figure} \ref{fig:KE-test}, we show the means and standard deviations of the right-hand side of \eqref{eq:tau_dependency2} averaged over $100$ realizations of $\chi_{1}^{(l)}(\omega_{\rm ex})$ for $\sigma=10^{-3}$ and $10^{-2}$.
For $(\tau,\alpha)=(2,0)$, the real and imaginary parts of the right-hand side of \eqref{eq:tau_dependency2} are wavy with period $\pi$ according to the choice of $\tau$, \blue{whereas} no wave is found for $(\tau,\alpha)=(0,2)$, as we expect from \eqref{eq:tau_dependency2}.
\blue{The large} standard deviations for large $|\omega_{\rm ex}-\Omega|$ are due to \blue{the} lack of population of oscillators, \blue{that is the} smallness of $g(\omega)$.
\blue{This} demonstration suggests that \blue{a} sufficiently precise observation of the susceptibility \blue{can identify} the existence and value of the time delay.
We \blue{must} remember that knowledge of the coupling strength and the natural frequency is assumed in the above demonstration\blue{; however,} it is usually difficult to access \blue{these} in advance.
Identification of the coupling strength and the natural frequency distribution is out of \blue{the} scope of this \blue{study} but \blue{should} be investigated.

\section{Summary and \blue{discussion}}\label{sec:summary}
We have developed a linear response theory for coupled oscillator systems \blue{by} directly solving the linearized equation of continuity around the nonsynchronized state.
One \blue{considerable} advantage of the proposed theory is that it is applicable to systems \blue{that have} general coupling functions, phase lags, and a time delay, while the previous two methods of the self-consistent equation and the Ott--Antonsen reduction are restricted \blue{to} the systems \blue{that have} a single sinusoidal coupling function.
\blue{The theoretical} predictions have been successfully verified by numerical simulations.

The reference state is assumed to be nonsynchronized as \blue{the} first step for a general theory.
\red{Another interesting topic would be}
to construct a linear response theory for other types of reference states: partially synchronized states, cluster states \cite{okuda-93}, chimera states \cite{kuramoto-battogtokh-02,abrams-mirollo-strogatz-wiley-08}, glassy states \cite{daido-92-2}, chaos states \cite{laing-12},  and so on.

Finally, we comment on a possible application of the linear response theory for the identification problem.
\blue{This} linear response theory gives macroscopic responses from microscopic details, and the identification problem can be formulated as the inverse problem.
A simple identification \blue{using} the proposed theory \blue{was} demonstrated in Section \ref{sec:daido_kuramoto} to distinguish between the phase lag and the time delay by surveying the dependence on the external frequency \blue{wtih} knowledge of the coupling strength and the natural frequency distribution.
A more systematic and precise identification method will be discussed elsewhere.

\ack
This work was supported by \blue{the Special Postdoctoral Research Program at RIKEN} (Y.T.) and JSPS KAKENHI Grand Numbers 19K20365 (Y.T.) and 16K05472 (Y.Y.Y.).

\appendix

\section{Analytic continuation}
\label{sec:continuation}
\blue{Function} $I_{n}(s)$ is \blue{first} defined in ${\rm Re}(s)>0$ in \eqref{eq:In}, which is the domain of the Laplace transform \eqref{eq:Laplace-transform-def}.
We continue this function into \red{the} \blue{entire} complex $s$ plane, which is necessary to obtain $I_{n}(in\omega_{\rm ex})$ included in the susceptibility $\chi_{n}(\omega_{\rm ex})$.
We have $I_{0}(s)=0$ and assume $n\neq 0$.

In the definition of $I_{n}(s)$ \eqref{eq:In}, the integral contour with respect to $\omega$ is the real axis and the contour does not meet the singular point $\omega=-is/n$ \red{because} $s$ is restricted in the right-half plane\red{,} ${\rm Re}(s)>0$.
\blue{However, in} the limit \blue{of} ${\rm Re}(s)\to +0$, the pole arrives on the real axis from the lower (upper) side of the complex $s$ plane for $n>0$ ($n<0$).
To avoid this pole, we smoothly modify the integral contour to the upper (lower) side and continue this modification for ${\rm Re}(s)<0$ so that we obtain the continued function.
This continuation gives the explicit form of function $I_{n}(s)$ as
\begin{equation}
  I_{n}(s)
  = \left\{
    \begin{array}{ll}
       \displaystyle\int_{-\infty}^{\infty}d\omega \displaystyle\frac{ing(\omega)}{s-in\omega} 
      \quad 
      &\mathrm{for}
      \quad
      {\rm Re}(s)>0, \\
      {\rm PV}  \displaystyle\int_{-\infty}^{\infty}d\omega \displaystyle\frac{ing(\omega)}{s-in\omega}
      \pm i\pi g\left(-\frac{is}{n}\right)
      \quad
      &\mathrm{for}
      \quad
      {\rm Re}(s)=0, \\
       \displaystyle\int_{-\infty}^{\infty}d\omega \displaystyle\frac{ing(\omega)}{s-in\omega}
      \pm 2 i\pi g\left(-\frac{is}{n}\right)
      \quad
      &\mathrm{for}
      \quad
      {\rm Re}(s)<0, \\
    \end{array}
  \right.
\end{equation}
where the second term for ${\rm Re}(s)\leq 0$ \blue{due to} the residue at the pole $\omega=-is/n$ and the positive (negative) sign corresponds to $n>0$ ($n<0$).

\section{Stability analysis \blue{using} the Nyquist diagram}
\label{sec:Nyquist}
The stability of the nonsynchronized state $F_{0}(\omega)$ is analyzed by turning off the external force, $H(\theta,t)=0$.
Roughly speaking, the initial perturbation\red{,} $f(\theta,\omega,0)$\red{,} plays the role of the external force in the response formula \eqref{eq:response-formula}\red{,} and the stability of $F_{0}(\omega)$ is determined by the zero points of $G_{n}(s)$, which is analytically continued \blue{via} the procedure \blue{in} \ref{sec:continuation}.
If there is a zero point whose real part is positive, then $F_{0}(\omega)$ is unstable.

\begin{figure}
  \centering
  \includegraphics[scale=0.60]{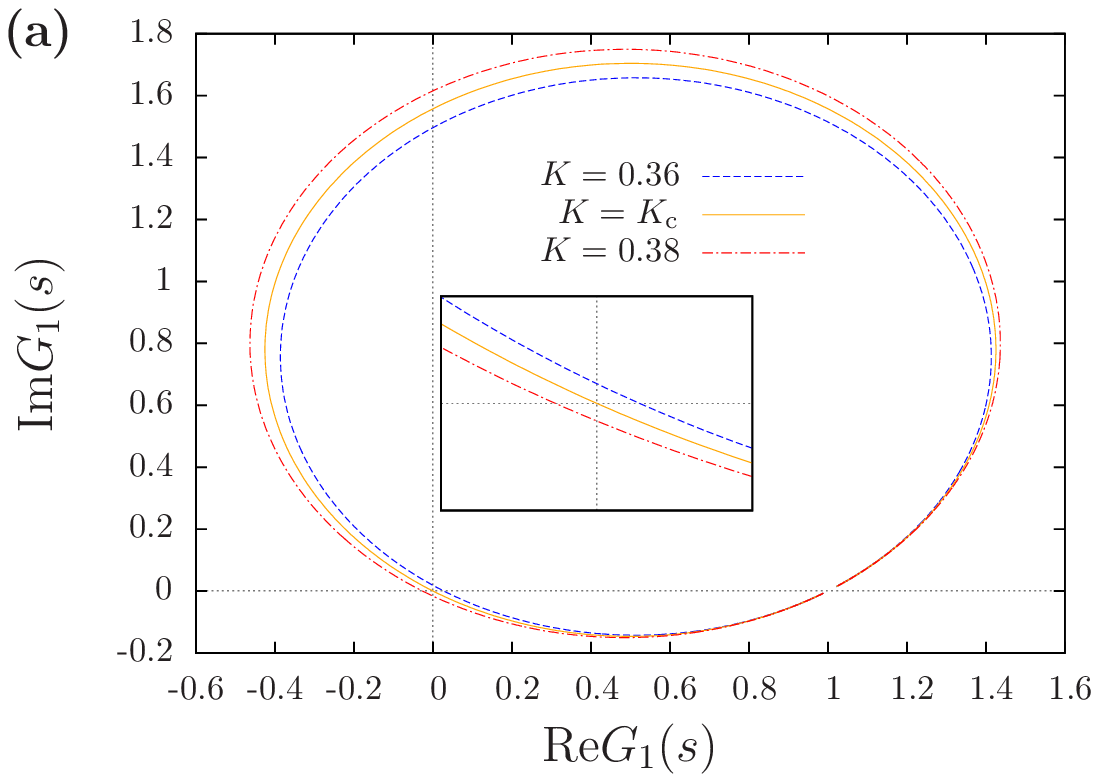}
  \includegraphics[scale=0.60]{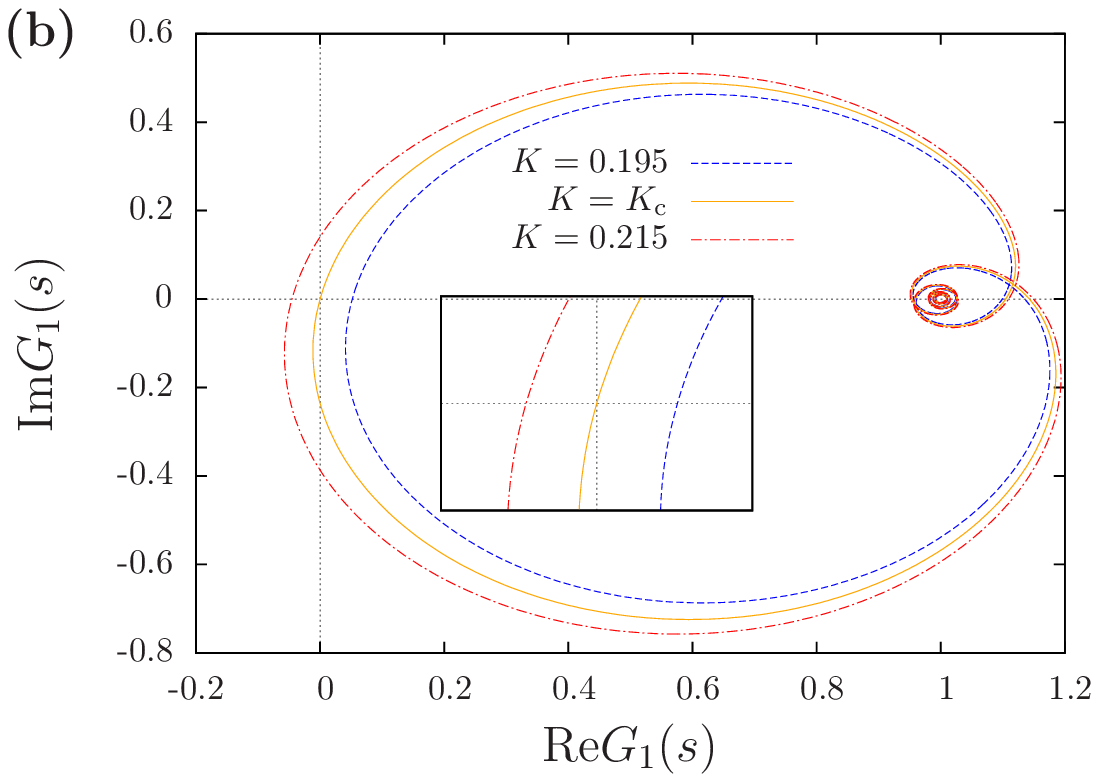}
  \caption{\blue{The} Nyquist diagrams.
    (a) The Sakaguchi--Kuramoto model. 
    $K_{1,\rm c}\simeq 0.370163$.
    (b) The time-delayed Daido--Kuramoto model.
    $K_{1,\rm c}\simeq 0.205665$.
    The natural frequency \red{distribution} $g(\omega)$
    is the single Lorentzian \eqref{eq:lorentz} with $(\gamma,\Omega)=(0.1,3)$.
    In each panel three curves are shown for $K<K_{\rm c}$ (blue broken),
    $K=K_{\rm c}$ (orange solid), and $K>K_{\rm c}$ (red dot-dashed).
    The inset is a magnification around the origin \red{of} the region \blue{of} $[-0.1,0.1]\times [-0.1,0.1]$.}
  \label{fig:Nyquist}
\end{figure}

We suppose that $G_{n}(s)$ is a mapping from the complex $s$ plane to the complex $G_{n}$ plane.
We focus on the boundary of the stability, the imaginary axis $s=iy~(y\in\mathbb{R})$.
In the limit of $y\to\pm\infty$ we have $G_{n}\to 1$.
\blue{Thus, the} mapped imaginary axis forms an oriented closed curve and the boundary of the unstable region ${\rm Re}(s)$ can be identified on the complex $G_{n}$ plane by the closed curve and its orientation.
This consideration implies that the nonsynchronized state\red{,} $F_{0}(\omega)$\red{,} is unstable if the mapped unstable region, \red{which is} the inside of the closed curve, contains the origin.
\blue{The} Nyquist diagrams for $G_{1}(s)$ are \blue{shown} in \blue{Figure} \ref{fig:Nyquist} for \blue{the} Sakaguchi--Kuramoto model and the time-delayed Daido--Kuramoto model.

\section{The Kuramoto model with an asymmetric natural frequency distribution}
\label{eq:Kuramoto-model}

We consider the Kuramoto model by setting the coupling function $\Gamma(\theta)$ as
\begin{equation}
  \Gamma(\theta) = - K \sin\theta
\end{equation}
and the parameters as $\alpha=\tau=0$ in \eqref{eq:Gamma-sin}.
The susceptibility for the order parameter\red{,} $z_{1}$\red{,} is written as
\begin{equation}
  \chi_{1}(\omega_{\rm ex})
  = \frac{\pi g(\omega_{\rm ex}) + i J(\omega_{\rm ex})}
  {2 - K [\pi g(\omega_{\rm ex}) + i J(\omega_{\rm ex})]},
  \label{eq:susceptibility_kuramoto}
\end{equation}
which is obtained through the general expression \eqref{eq:chin}.

We employ an asymmetric natural frequency distribution\red{,} $g(\omega)$\red{,} to observe \blue{the} non-divergence of the susceptibility at the critical point for the static external force and show \blue{the} qualitative difference from the other two types of asymmetry, the phase \blue{lag} and the time delay.
The asymmetry is \red{achieved} by the family of $g(\omega)$ as
\begin{equation}
  g(\omega) = \frac{c}{\left[(\omega-\Omega)^{2}+\gamma_{1}^{2}\right]\left[(\omega+\Omega)^{2}+\gamma_{2}^{2}\right]},
  \label{eq:product_lorentz}
\end{equation}
where $\Omega\geq 0, \gamma_{1},\gamma_{2}>0$ and the normalization
constant $c$ is given by
\begin{equation}
  c = \frac{\gamma_{1}\gamma_{2}[ (\gamma_{1}+\gamma_{2})^{2} + 4\Omega^{2}]}
  {\pi (\gamma_{1}+\gamma_{2})}
\end{equation}
as in \blue{ref.} \cite{terada-ito-aoyagi-yamaguchi-17}.
The distribution is symmetric if $\gamma_{1}=1$ or $\Omega=0$ and tends to be bimodal with large $\Omega$.
We use the parameter set  $(\gamma_{1},\gamma_2,\Omega)=(0.6,1,0.6)$ \blue{to realize the} asymmetric unimodal\red{,} $g(\omega)$.

\begin{figure}
  \centering
 \includegraphics[scale=0.65]{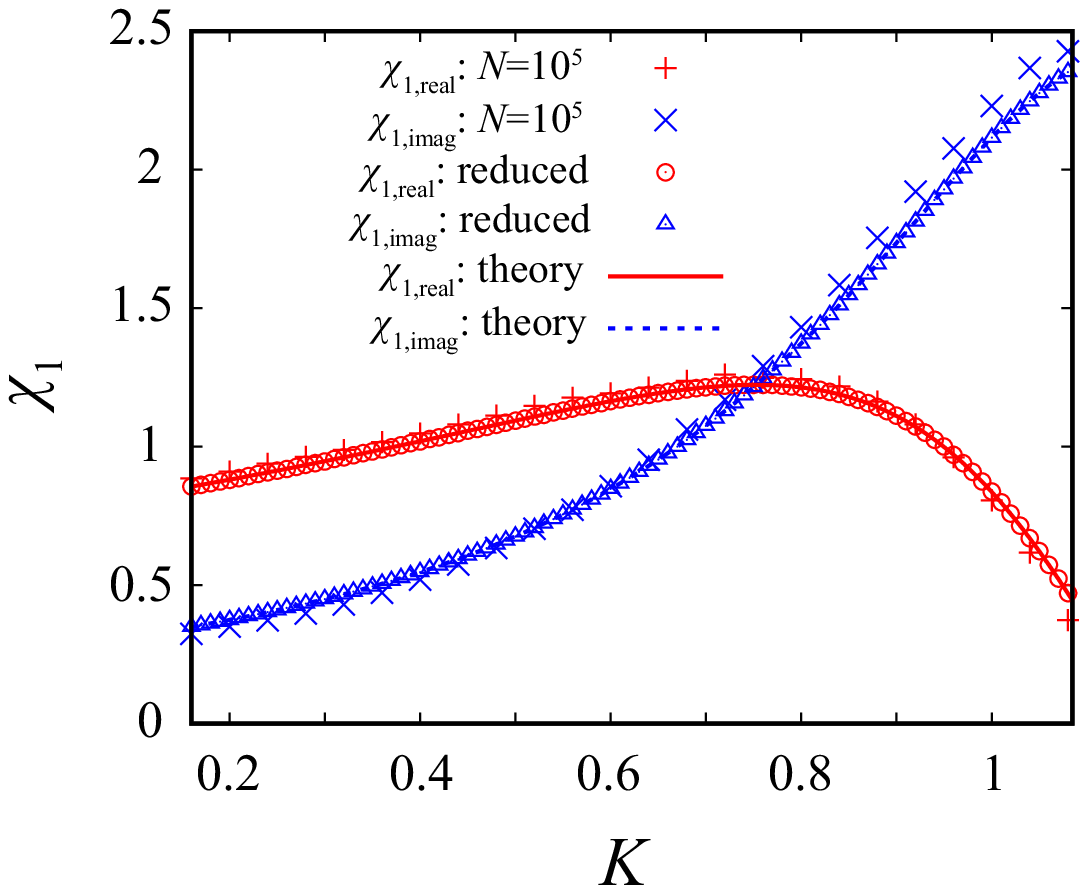}
  \caption{Susceptibility in the Kuramoto model with the asymmetric natural frequency distribution, where $(\gamma_{1},\gamma_2,\Omega)=(0.6,1,0.6)$ in \eqref{eq:product_lorentz}.
   The frequency of the external force is set to zero: $\omega_{\rm ex}=0$.
   There is no divergence at the critical point $K=K_{\rm c}\simeq 1.084618$, which is the right boundary of the panel.
   \blue{Numerical} simulations are conducted with $N=10^{5}$ and $h=10^{-2}$. }
  \label{fig:susceptibility_cor_sym}
\end{figure}

\blue{First}, we consider the static external force case with $\omega_{\rm ex}=0$.
The susceptibility does not diverge even at the critical point\red{,} $K_{\rm c}$\red{,} as shown in \blue{Figure} \ref{fig:susceptibility_cor_sym}.
This nondivergence is in good agreement with the first remark mentioned in \blue{Section} \ref{sec:susceptibilities}.

\begin{figure}
  \centering
  \includegraphics[scale=0.65]{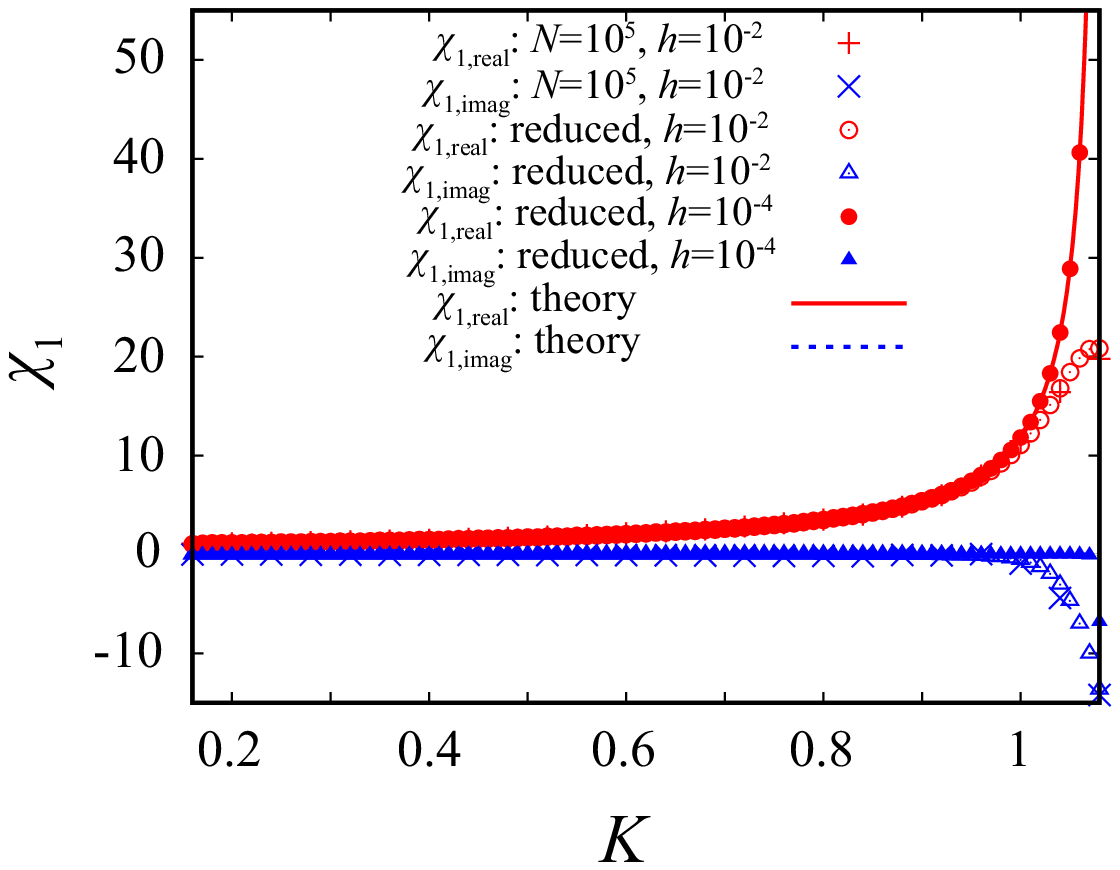}
  \caption{Susceptibility in the Kuramoto model  with the asymmetric natural frequency distribution,  $(\gamma_{1},\gamma_2,\Omega)=(0.6,1,0.6)$ in \eqref{eq:product_lorentz}.
    The frequency of the external force is set \red{to be} $\omega_{\rm ex}=0.303819$, which satisfies $J(\omega_{\rm ex})=0$.
    The divergence of $\chi(\omega_{\rm ex})$ is observed at the critical point\red{,} $K=K_{\rm c}\simeq 1.084618$, \blue{which is} the right boundary of the panel.
    The numerical simulations are conducted with $N=10^{5}$ and $h=10^{-2},10^{-4}$.
}
  \label{fig:susceptibility_cor_sym_omega_ex}
\end{figure}

The divergence \blue{recovers} if we choose the external frequency\red{,} $\omega_{\rm ex}$\red{,} \blue{appropriately}.
The divergence appears under the condition $J(\omega_{\rm ex})=0$ at the point \red{of}
\begin{equation}
  K_{\rm c} = \frac{2}{\pi g(\omega_{\rm ex})}.
  \label{eq:Kc-K-omegaex}
\end{equation}
This theoretical prediction is confirmed numerically in \blue{Figure} \ref{fig:susceptibility_cor_sym_omega_ex}\blue{, where we can observe their discrepancy with larger $h$ because of the nonlinearity of the response as explained in Section \ref{sec:sk_model}.}

In the Kuramoto model, the divergence of \blue{the} susceptibility appears only in the real part, which \red{is shown in} Eq. \eqref{eq:susceptibility_kuramoto}.
In other words, the response must be parallel to the direction of the external force.
\blue{In contrast}, in the Sakaguichi--Kuramoto model and the time-delayed Daido--Kuramoto model, the phase lag and the time delay permit divergences in both real and imaginary parts of the susceptibility, and the direction of \blue{the} response is not always parallel to the external force, \blue{\textit{i.e.}}, we may observe a phase-gap between the external force and the linear response.
The coexistence of the divergence and the phase gap thus reveals \blue{the} different roles of the two groups of asymmetry; asymmetry by the phase lag or the time delay permits the coexistence, while asymmetry in $g(\omega)$ does not.

\section{Calculation of susceptibility with Lorentzian natural frequency distributions}
\label{sec:lorentz_case}

In our numerical simulations, we use a Lorentzian or \red{its} product as the natural frequency distributions in the main text.
In that case the principal value\red{,} $J(\omega_{\rm ex})$ \eqref{eq:definition_J}, can be computed explicitly by using the residue theorem.
We introduce the integral
\begin{equation}
  \label{eq:Jepsilon-def}
  J_{\epsilon}(\omega_{\rm ex}) = \int_{-\infty}^{\infty} d\omega
  \frac{g(\omega)}{\omega-\omega_{\rm ex}+i\epsilon}.
\end{equation}
The pole is at $\omega=\omega_{\rm ex}-i\epsilon$ and the integral is well defined for $\epsilon\neq 0$.
We assume $\epsilon>0$.
Moreover, we consider the single Lorentzian \eqref{eq:lorentz} for simplicity.
Adding the upper half-circle contour and using the residue theorem \blue{by} picking up the pole $\omega=\Omega+i\gamma$,
we \red{are able to} compute the value of $J_{\epsilon}(\omega_{\rm ex})$ explicitly as
\begin{equation}
  \label{eq:Jepsilon}
  J_{\epsilon}(\omega_{\rm ex})
  = \frac{1}{\Omega-\omega_{\rm ex}+i(\gamma+\epsilon)}.
\end{equation}
Taking the limit $\epsilon\to +0$ in \eqref{eq:Jepsilon} and applying the continuation technique presented in \ref{sec:continuation} to the right-hand side of \eqref{eq:Jepsilon-def}, we have
\begin{equation}
  \lim_{\epsilon\to +0} J_{\epsilon}(\omega_{\rm ex})
  = J(\omega_{\rm ex}) - i\pi g(\omega_{\rm ex})
\end{equation}
and
\begin{equation}
  J(\omega_{\rm ex})
  = \frac{1}{\Omega-\omega_{\rm ex}+i\gamma} + i\pi g(\omega_{\rm ex})
  = \frac{\Omega-\omega_{\rm ex}}{(\Omega-\omega_{\rm ex})^{2}+\gamma^{2}}.
\end{equation}
The above idea is applicable to the multiplicative Lorentzian \eqref{eq:product_lorentz}.

\section{Ott--Antonsen reduction in \blue{the} case with Lorentzian natural frequency distributions}
\label{sec:oa_reduction}
In the numerical confirmation of the Sakaguchi--Kuramoto model, we used the Ott--Antonsen ansatz \cite{ott-antonsen-08,ott-antonsen-09}, which reduces the equation of continuity to a real two-dimensional system describing \blue{the} dynamics of the order parameter $z_{1}$.
This reduction is limited to a class of models, but the reduced equations are useful \blue{for examinining} our theory because the reduced system corresponds exactly to the large population limit and has no finite-size fluctuations.
\blue{Here, we} derive the reduced equations in the Sakaguchi--Kuramoto model with the natural frequency \eqref{eq:product_lorentz}.

The Ott--Antonsen ansatz introduces the form of the probability density function $F$ as
\begin{equation}
  \label{eq:oa_ansatz}
  F(\theta,\omega,t)
  = \frac{g(\omega)}{2\pi} 
  \left\{ 1 + \sum_{n=1}^{\infty} \left[
      a^{n}(\omega,t) e^{in\theta}
    \right. \right. 
  + \left. (a^{\ast}(\omega,t))^{n} e^{-in\theta}
  \right] \biggr\},
\end{equation}
where the complex-valued function $a(\omega,t)$ satisfies the condition $|a\left(\omega,t\right)|<1$ and is regular on the complex $\omega$-plane.
Substituting the ansatz \eqref{eq:oa_ansatz} into the equation of continuity for the Kuramoto model with the first harmonic external force having the frequency \red{of} $\omega_{\mathrm{ex}}$, we obtain the equation for $a\left(\omega,t\right)$ as
\begin{equation}
  \label{eq:oa_a_z}
  \frac{\partial a}{\partial t} = -i\omega a+\frac{K}{2}\left(z_1^{\ast}-a^2z_1\right)
  -\frac{h}{2}\left(e^{-i\omega_{\mathrm{ex}}t}-a^2e^{i\omega_{\mathrm{ex}}t}\right),
\end{equation}
where the order parameter $z_1(t)$ and $a(\omega,t)$ are related through
\begin{equation}
  z_1 = 
  \int_{-\infty}^{\infty} d\omega\, g(\omega) a^{\ast}(\omega,t),
  \label{eq:relation_a_z}
\end{equation}
which is obtained from \eqref{eq:oa_ansatz}.
The integral over $\omega$ in the right-hand side of \eqref{eq:relation_a_z} is performed by adding the large upper-half circle, which has no contribution to the integral, and picking up the two poles of $g(\omega)$, \eqref{eq:product_lorentz}, at $\omega=\Omega+i\gamma_{1}$ and $\omega=-\Omega+i\gamma_{2}$.
The residues give
\begin{equation}
  z_1(t) = k_{1} A(t) + k_{2} B(t),
\end{equation}
where complex variables $A$ and $B$ are defined by
\begin{equation}
  A(t) = a^{\ast}(\Omega+i\gamma_{1},t) ,\quad
  B(t) = a^{\ast}(-\Omega+i\gamma_{2},t) 
\end{equation}
and the time-independent coefficients are given by
\begin{equation}
    k_{1}  = \frac{\gamma_{2}}{\gamma_{1}+\gamma_{2}}
    \frac{2\Omega-i\left(\gamma_1+\gamma_2\right)}
    {2\Omega+i\left(\gamma_1-\gamma_2\right)},
    \quad
    k_{2}  = \frac{\gamma_{1}}{\gamma_{1}+\gamma_{2}}
    \frac{2\Omega+i\left(\gamma_1+\gamma_2\right)}
    {2\Omega+i\left(\gamma_1-\gamma_2\right)}.
\end{equation}
Finally, in \eqref{eq:oa_a_z}, \blue{by} setting $\omega$ as $\omega=\Omega+i\gamma_{1}$ or $\omega=-\Omega+i\gamma_{2}$, we have the reduced equations
\begin{eqnarray}
\frac{dA}{d t} &= 
i\left(\Omega+i\gamma_1\right)A
-\frac{K}{2}\left[A^2\left(k^{\ast}_1 A^{\ast} + k^{\ast}_2 B^{\ast} \right)e^{i\alpha}-\left(k_1A+k_2B\right)e^{-i\alpha}\right]\nonumber\\
&\quad-\frac{h}{2}\left(A^2e^{-i\omega_{\mathrm{ex}}t}-e^{i\omega_{\mathrm{ex}}t}\right),\label{eq:reduced_eq1}\\
\frac{dB}{d t} &= i\left(-\Omega+i\gamma_2\right)B
-\frac{K}{2}\left[B^2\left(k^{\ast}_1 A^{\ast}+k^{\ast}_2 B^{\ast} \right)e^{i\alpha}-\left(k_1A+k_2B\right)e^{-i\alpha}\right]\nonumber\\
&\quad-\frac{h}{2}\left(B^2e^{-i\omega_{\mathrm{ex}}t}-e^{i\omega_{\mathrm{ex}}t}\right).\label{eq:reduced_eq2}
\end{eqnarray}

Similarly, for the Sakaguchi--Kuramoto model with the Lorentzian $g(\omega)$, which is \blue{discussed} in \blue{Section} \ref{sec:sk_model}, we have the reduced equation for $z_1$:
\begin{equation}
  \frac{dz_1}{dt}
  = 
    i (\Omega + i\gamma) z_{1}
    - \frac{K}{2}z_1\left( |z_1|^2e^{i\alpha}- e^{-i\alpha} \right)
    - \frac{h}{2}\left(z_1^2 e^{-i\omega_{\mathrm{ex}}t}- e^{i\omega_{\mathrm{ex}}t} \right).
\end{equation}
It is worth noting that the Ott--Antonsen reduction is also applicable to the time-delayed Kuramoto model \cite{ott-antonsen-08}.

\end{document}